\documentclass[10pt]{article}

\usepackage[normalem]{ulem}
\usepackage{fullpage}

\usepackage{textcomp}
\usepackage{amssymb}
\usepackage{booktabs}
\usepackage{float}
\usepackage{times}
\usepackage{mdframed}
\usepackage{latexsym}
\usepackage{graphicx}
\usepackage{tabularx}
\usepackage{wrapfig}
\usepackage{url}
\usepackage{multirow}
\usepackage{hyperref}
\usepackage{framed}
\usepackage{color}
\usepackage{xcolor}
\usepackage{xspace}
\usepackage{amsmath}
\usepackage{amssymb}
\usepackage{arydshln}
\usepackage{makecell}
\usepackage{amsmath}
\usepackage{xcolor}

\newcommand{\bugid}[1]{#1}

\usepackage{listings}

\definecolor{mGreen}{rgb}{0,0.6,0}
\definecolor{mGray}{rgb}{0.5,0.5,0.5}
\definecolor{mPurple}{rgb}{0.58,0,0.82}
\definecolor{backgroundColour}{rgb}{0.94, 0.97, 1.0}

\lstdefinestyle{CStyle}{
    backgroundcolor=\color{backgroundColour},
    commentstyle=\color{mGreen},
    keywordstyle=\color{magenta},
    numberstyle=\tiny\color{mGray},
    stringstyle=\color{mPurple},
    breakatwhitespace=false,
    breaklines=true,
     basicstyle=\scriptsize\ttfamily,  
    captionpos=b,
    keepspaces=true,
    numbers=left,
    numbersep=4pt,
    showspaces=false,
    showstringspaces=false,
    showtabs=false,
    tabsize=2,
    language=C
}

\newcommand{\Og}{-{\it Og}\xspace}

\newcommand{\Otre}{-{\it O3}\xspace}

\newcommand{\optocus}{Debug$^2$\xspace}
\newcommand{\n}{\optocus}
\newcommand{\code}{\texttt}
\newcommand{\cd}{\code}

\newcommand{\newpara}[1]{\smallskip\noindent\textit{#1}~\textendash~}
\newcommand{\newparabold}[1]{\medskip\noindent\textbf{#1}~\textendash~}

\newcommand{\aurl}[1]{\url{#1}}
\begin{document}

\title{Who's Debugging the Debuggers?\\\vspace{0.5em}\large Exposing Debug Information Bugs in Optimized Binaries}

\author{Giuseppe Antonio Di Luna$^{1}$\footnote{Supported by the AXA Postdoctoral Fellowship}, Davide Italiano $^{2}$, Luca Massarelli$^{1}$,\\ Sebastian \"Osterlund$^{3}$, Cristiano Giuffrida$^{3}$, Leonardo Querzoni$^{1}$ \\
	{1}: Sapienza, University; \\
	{2}: Apple;  \\
	{3}: Vrije Universiteit Amsterdam.\\
}
\date{}
\maketitle


\begin{abstract}
Despite the advancements in software testing, bugs
still plague deployed software and result in crashes 
 in production. When debugging issues ---sometimes caused by ``heisenbugs''--- there is the need to interpret
core dumps and reproduce the issue offline on the \emph{same} binary deployed. This requires the entire toolchain (compiler, 
linker, debugger) to correctly generate and use debug information. 
Little attention has been devoted to
checking that such information is correctly preserved by modern toolchains'
 optimization stages. This is particularly important as managing debug
information in optimized production binaries is non-trivial, often
leading to toolchain bugs that may hinder post-deployment debugging efforts.

In this paper, we present \n, a framework to find debug information bugs
in modern toolchains. Our framework feeds random source programs
to the target toolchain and surgically compares the debugging behavior of their
optimized/unoptimized binary variants. Such differential
analysis allows \n to check invariants at each debugging step and
detect bugs from invariant violations. Our invariants are based on the
(in)consistency of common debug entities, such as source lines,
stack frames, and function arguments. We show that, while simple, this strategy
yields powerful cross-toolchain and cross-language invariants, which can
pinpoint several bugs in modern toolchains. We have used \n to find
23 bugs in the LLVM toolchain (clang/lldb), 8 bugs in the GNU toolchain
(GCC/gdb), and 3 in the Rust toolchain (rustc/lldb)---with 14 bugs already
fixed by the developers.
\end{abstract}

\section{Introduction}

Production binaries need to be heavily optimized to maximize metrics such as speed, size, and energy consumption~\cite{schulte2014post}. For this purpose, modern compilers feature optimization stages where several sophisticated transformation passes cooperate to produce the final binary. Preserving debug information in this process is important for several reasons.

First, correct debug information helps the interpretation of core dumps collected in production (e.g., to provide line information in stacks leading to a crash). More importantly, correct debug information is crucial when reproducing and debugging production issues offline on the \emph{same} optimized binary. This is often necessary since compiler optimizations also alter the observability of unwanted behaviors compared to the unoptimized case. In other words, common issues such as race conditions~\cite{6595791}, memory errors~\cite{7163211}, and other classes of ``heisenbugs''~\cite{YIN201918} may not even be reproducible without a reliable debugging process for optimized binaries~\cite{hennessy1982symbolic}.

However, preserving debug information for optimized production binaries is a daunting task, with no obvious mapping between source and assembly statements~\cite{copperman1994debugging,hennessy1982symbolic} and the potential to introduce bugs at each of the several layers of modern toolchains. As we will show, efforts to provide debug-friendly optimization levels (such as -$Og$ in modern compilers) fall short on providing a bug-free debugging experience.

Yet, despite the challenges and relevance of this task, as well as toolchain developers' efforts to improve the debuggability of optimized binaries, little attention has been devoted to scrutinizing the full debug information lifecycle for bugs. Prior work largely focused on debugger testing~\cite{tolksdorf2019interactive, lehmann2018feedback}, with one recent exception focusing on the ability to retrieve correct variable values in optimized binaries~\cite{davidepldi}. We stress that the \emph{entire} toolchain (compiler, linker, debugger) must be free of debug information bugs to provide a reliable debugging experience.

In this paper, we introduce \n, a framework to expose debug information bugs in production toolchains. The key idea is to feed random source programs to a target toolchain and compare the debugging behavior of their optimized/unoptimized binary variants to expose bugs. To achieve this, \n first extracts {\em debugging traces} of each binary by single-stepping its execution in the target debugger. Next, \n performs differential analysis of each pair of optimized/unoptimized traces to check for unexpected differences at each step. To check for (un)expected behavioral differences, \n relies on \emph{trace invariants} empirically based on the (in)consistency of common debug elements, such as source lines, stack frames, and function arguments.
Our trace invariants are explicitly designed to be generic (i.e., toolchain- and programming language-agnostic) and, while simple, can effectively pinpoint inconsistencies in the entire debug information lifecycle. As we will show, these inconsistencies can then be traced back to the underlying issue, exposing  bugs in the entire toolchain; including all components of the compiler (from backend until the optimization stages) and the debugger. 

Our analysis shows that after decades of development, mature toolchains still suffer from a conspicuous amount of bugs that \n can automatically find.  Surprisingly enough, many of these bugs plague the optimization levels specifically created for a smooth debug experience (i.e., -$Og$).

\newparabold{A motivating example} Snippet~\ref{mexampleclang} shows a bug \n exposed in the LLVM toolchain when compiling with -$Og$.

\n can expose this bug with a simple \emph{source line invariant}: a line containing dead code for a given binary should be absent from its debugging trace. However, while analyzing the trace, \n detects the debugger eventually pointing to the dead line 8 --- which is never executed, as opposed to line 7 --- and flags an invariant violation. While \n detects this issue by line stepping in the debugger (which previous work instead assumed to be correct~\cite{davidepldi}), this is actually a compiler bug. In particular, the bug is caused by the compiler backend's branch folding pass and it is part of a more general class of bugs in which optimization passes move instructions across basic blocks without properly updating the corresponding debug information. While these bugs clearly degrade the debugging experience by displaying misleading execution flows, they still escape state-of-the-art testing efforts and are surprisingly common.

\begin{wrapfigure}[12]{r}{0.21\textwidth}

  \begin{lstlisting}[caption={Clang bug \bugid{46009}: wrong step at line 8.},captionpos=b,style=CStyle,label=mexampleclang]
int a, b, c;
int main()
{
  {int ui1 = 5, ui2 = b;
    c =
      ui2 == 0 ?
      ui1 :
      (ui1 / ui2);
  }
}
\end{lstlisting}
\end{wrapfigure}

We have used \n to expose many such toolchain bugs, in passes ranging from control-flow graph simplification to loop-invariant code motion. The reported issues sparked lively discussions among toolchain developers, evidencing the lack of much needed guidelines defining how to preserve debug information during optimization passes. This forces developers to find the best course of action on an issue-by-issue basis, slowing down the rate at which bugs can be fixed. Interestingly, our bug reports influenced an initial document published by LLVM developers on how to handle debug information during optimization passes \cite{vedantkumarDBG}.

We hope our work will serve as an inspiration to evolve the standard UNIX debugging format (DWARF), which currently lacks proper support to represent the effect of transformations on the source-to-binary mapping. Even simple optimizations (e.g., common subexpression elimination) struggle to correctly preserve debug information due to DWARF's inability to map a single address to multiple source locations.

\medskip\noindent\textbf{Contributions}~\textendash~
We make the following contributions:
\begin{itemize}
\item \n, a framework to scrutinize the debug information lifecycle for optimized binaries using trace invariants (\S\ref{system:overview}).
\item Four trace invariants to pinpoint debugging behavioral differences in optimized/unoptimized binaries and expose toolchain bugs. \n's invariants are empirically derived and designed to expose bugs across different toolchains and programming languages (\S\ref{sec:generalidea}).
\item An extensive experimental evaluation of \n on the LLVM toolchain. We also present experiments on the GNU and Rust toolchains to confirm the generality of our approach. \n exposed bugs in all such toolchains (\S\ref{sec:eval}), specifically 23 bugs in the LLVM toolchain (clang/lldb), 8 bugs in the GNU toolchain (GCC/gdb), and 3 in the Rust toolchain (rustc/lldb). The developers have already confirmed 22 of these bugs, and 14 of these have been fixed.
\item Lessons learned from our interactions with the toolchain developers, egregious bugs found, and current shortcomings in the DWARF debugging format  (\S\ref{sec:paperdiscussion} and \S\ref{sec:bugdiscussion}). 
\end{itemize}

\section{The \optocus Framework}\label{system:overview}

\begin{figure*}[h]
	\centering
	\includegraphics[width=\textwidth]{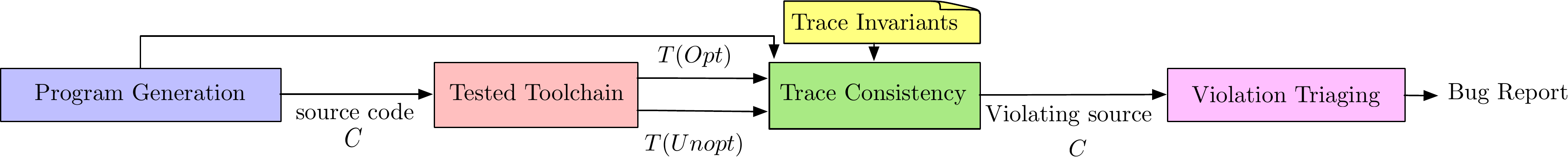}
	\caption{Framework Overview}\label{fig:framework_overview}
\end{figure*}

The \n framework is based on a simple idea: consider a software that is compiled and debugged with a given toolchain; the execution traces obtained from the debugger will, in general, differ if they are obtained from versions compiled with different optimization flags, but their semantics must match. In particular, there is some behavior that must manifest coherently in all traces, no matter the optimization level used in the compilation process. For example, a line in the source that represents dead code must be absent from any execution trace.

Our framework uses a set of these intuitive rules, that we call \emph{trace invariants}, to check if the traces of differently optimized binaries are coherent or not. Whenever an invariant is violated, an inconsistency is detected that points to a candidate bug in the toolchain. Candidates are then verified before reporting them to developers.

\n encompasses four main modules (see Figure \ref{fig:framework_overview}):

\begin{itemize}
	\item {\bf Program Generation}: responsible for generating the source code samples used to test a specific toolchain; the output of this module is the source code $C$.
	\item {\bf Tested Toolchain}: the toolchain that is being tested that typically includes at least a compiler and a debugger; the compiler is used to compile $C$ using both a desired optimization level, and with no optimization; the two resulting binaries are then fed to the debugger that outputs the two traces $T(opt)$ and $T(unopt)$.
	\item {\bf Trace Consistency}: takes as input the two traces and checks if any of the trace invariants are violated, indicating an inconsistency that could point to a bug in the toolchain producing the optimized trace. 
	This module outputs any source code $C$ that violates at least one trace invariant.

\item {\bf Violation Triaging}: this module reduces the large number of violations caused by the same candidate bug by clustering them and extracts a fragment of representative code for each cluster. This simplifies the manual analysis of the candidates to remove false positives and report to developers only highly probable bugs.
\end{itemize}

\subsection{Program generation} \label{sec:codegen}

The program generation module has the sole purpose of generating source code to be fed to the tested toolchain module. The only requirement for this module is that it must output a source code $C$ that can be correctly compiled by the tested toolchain and run; moreover, to simplify the design of the trace invariants, we assume the generated programs to be well-formed, deterministic, closed (i.e., don't take any input), and single-threaded.

There are several strategies that can be used when implementing this module. We can follow a purely generative approach, e.g., by using tools like Csmith \cite{yang2011finding} or Yarpgen \cite{yarpgen}; or a more refined mutational approache guided by some feedback loop (e.g., using the coverage of the tested toolchain).
In our implementation of the framework, we mainly used the generative approach (leveraging both Csmith and Yarpgen); we also performed an analysis implementing a guided mutational approach that prefers mutated samples increasing the internal coverage of the tested toolchain. However, we did not find it to bring improvements over the pure generative approach.

Details are reported in \S\ref{sec:codegeneval}. 
\vspace{-0.2cm}

\subsection{Tested toolchain} 
\label{sec:tested_toolchain}
\vspace{-0.2cm}
This module integrates the toolchain to be tested by \n, it takes as input the source code $C$ and it outputs the two traces \emph{T(opt)} and \emph{T(unopt)} that will be analyzed by the trace consistency module.
More precisely, we model the toolchain's compiler as an object that takes an input a source code $C$, and outputs a compiled program $P$. We assume that $P$ contains debug information (practically speaking $P$ has been compiled with -$g$ flag).  $P$'s code can be either optimized or not, depending on the optimization flags used to customize the compiling phase.
Since we are interested in comparing the debug information of an optimized and unoptimized binary generated from the same code $C$,  we will indicate with \emph{unopt} the unoptimized version of $P$, and with \emph{opt} the optimized counterpart.

Using the toolchain's debugger, we execute $P$ repeatedly stepping over source lines until it exits, obtaining an {\em execution trace} $T(P)$ (i.e., an ordered list of elements, each being a step over a source line). We assume the following information are associated to a step $s \in T(P)$:

\begin{itemize}
	\item \textbf{Line information} -- Given $s$, the \emph{line} function $l(s)$ returns either a line in the source code $C$ or a dummy line $\bot$. In practice, this information is  provided by the majority of debuggers on executables compiled with debug information.
%
The dummy line $\bot$ models the absence of line information caused by its dropping in some optimization passes; for example, \emph{clang} sometimes creates line information that points to the artificial line $0$, which is the equivalent of our $\bot$.


	\item \textbf{Variables information} -- Given $s$, the \emph{variables} function $V(s)$ returns the set of all global variables, local variables, and parameters visible at step $s$. Given a variable $v$ in $V(s)$ we use $value(v,s)$ to indicate its value at step $s$. We will use the special value $\bot$ to model optimized out variables; i.e., variables that have been removed by some optimization passes while producing $P$. Function $V$ can be for example implemented in lldb with command \cd{frame var}.

	\item \textbf{Backtrace information}: Given $s$, the \emph{backtrace} function $BT(s)$ returns the set of function names present in the stack before the execution of $s$ (n.b., usually this set is ordered, but in our paper, we never make use of such an ordering). An example is the output of the \cd{bt} command in lldb or gdb.

\end{itemize}
This set of information is the one that will be used to define the four trace invariants presented in \S\ref{sec:generalidea}. However, the framework may accommodate a richer definition of trace based on further information available from the debugger. 
Our trace invariants assume that, at each step, one of the traces contains correct information.  Another assumption that we do is that the unoptimized trace steps on all the source-lines that are executed, we found this assumption to be true since the compiler, with all optimization turned off, lowers any executable source code line to at least one assembly instruction.

\subsection{Trace consistency}

The trace consistency module analyzes the \emph{T(unopt)} and \emph{T(opt)} traces to find inconsistencies that are likely due to a bug in the toolchain (usually in the compiler or in the debugger).
The module takes as input two debug traces \emph{T(unopt)} and \emph{T(opt)} and a set of trace invariants.
Each invariant takes the two traces \emph{T(Opt), T(Unopt)}, analyses the information contained, and raises a {\em violation} in case an inconsistency is found. For each violation, the corresponding source code C is sent to the triaging module. 
In this paper, we present four trace invariants (see \S \ref{sec:generalidea}), but this module can easily accommodate further invariants that reason on the trace contents.

 
\subsection{Violations triaging}
The last module of \n is used to cluster test cases exhibiting violations that are likely to share the same ``root cause'', i.e., that are caused by the candidate bug. 
This is important to reduce the number of candidate bugs to be verified, as this verification must be performed manually.

Once we find that a certain source code $C$ generates a violation on the toolchain under test, we
extract a {\em fingerprint} by testing $C$ against a set of different versions of the same toolchain (e.g., using several releases of the same compiler).
Each of these versions applies a different sequence of optimization passes. We progressively apply the passes in the sequence until we find the first pass that causes the violation; the name of this pass will be included in the fingerprint for the specific toolchain version under analysis; if no violation happens, we include a default value in the fingerprint.
Given a source code $C$, we then obtain a vector $F(C)$, where each component corresponds to the output of our procedure on a specific version of the toolchain under test. This vector is the {\em fingerprint} of $C$. 

For example, consider a source code sample $C$ that violates a trace invariant on a given toolchain $X$. To build its fingerprint, we test the same code on $n$ different versions of $X$, (e.g., 5.0, 6.0, 7.0, 8.0 and 9.0), and check for each version which optimization pass first violates the invariant. Assume the violation happens on pass $j$ for versions 5.0 and 6.0, while it happens on pass $k$ for the other versions. Then the fingerprint for $C$ will be \emph{F(C)=[``j'', ``j'', ``k'', ``k'', ``k'']}

Note that only using toolchain versions in the fingerprint would not permit to discriminate between two bugs that appear in the exact same versions but are in two different optimization passes. For a similar reason, we cannot use a single toolchain version in the fingerprint.  We cluster together source samples with the same fingerprint as they link to a single candidate bug with high probability. Then, we select a random sample for each cluster and extract from it a piece of minimal code that allows an easier manual analysis.

\section{Trace Invariants}\label{sec:generalidea}


In this paper, we introduce four trace invariants, designed to detect different problems in a target trace. They are:
 \begin{mdframed}[backgroundcolor=blue!8]

\begin{description}
\item[The Line Invariant (LI)] checks for misstepped lines;
\item[The Backtrace Invariant (BI)]  checks for spurious frames in the backtrace of a line;
\item[The Scope Invariant (SI)] checks if there are out-of-scope variables;
\item[The Parameters Invariants (PI)] checks the consistency of the values assumed by function parameters. 
\end{description}
   \end{mdframed}

Note that our set of four trace invariants does not pretend to be complete; that is, further invariants may be defined that could possibly allow the identification of further bugs whose semantics is not captured by the invariants presented here.

For each of our invariants, we introduce it and justify the assumption behind its design; then, we give a more formal definition and present an example of a real-world bug discovered with it.  

\newparabold{Lines Invariant (LI)}
Lines Invariant checks whether there exists a line of $C$ that appears to be executed in the optimized trace while the unoptimized trace never executes it.
The rationale behind LI is that such a spurious line is either dead code or a glitch of the compiler/debugger (e.g., stepping over a variable declaration); this comes directly from the fact that $C$ is closed and deterministic, see \S\ref{sec:codegen} and that the unoptimized trace steps on all source lines that are executed.

\newpara{Formal definition} Given a program $P$ we indicate with $L(P)$ all the lines in $T(P)$ (i.e., $L(P):\{l(s) | s \in T(P)\}$). The line subset invariant is violated when $L(opt) \not\subseteq L(unopt) \cup \{\bot\}$. Note that the definition excludes the case in which the line where the violation happens is $l(s)=\bot$; this was done to model corner cases that arise when using our invariant on specific instances of programming languages, compilers, and debuggers (e.g., the aforementioned line $0$ in clang).

\newpara{Violation of LI} Snippet \ref{lis} shows an example of a violation of LI in C code compiled with clang and optimization -$Og$. In this case, we have $4 \in L(opt)$ and $4 \not\in L(unopt)$; line 4 is clearly dead code, and stepping on it is a bug caused by wrong debug information.

\noindent \begin{minipage}{.5\textwidth}
   \begin{lstlisting}[caption={LI violation.  Line 4 appears to be executed.},captionpos=b,style=CStyle,label=lis]
int a;
int b(char c) {
  return (a>=2||c>>a)
             ? c
             : 0;
}
int main(){return b(0);}
\end{lstlisting}
\end{minipage}
\,\,\,\,
\begin{minipage}{.5\textwidth}
   \begin{lstlisting}[caption={BI violation. Backtrace at line 8 contains \cd{func\_2}.},captionpos=b,style=CStyle,label=snippet1]
static int a;
static int *b = &a;
static int *func_2() {
  *b = 0;
  return &a;
}
void func_1() { func_2(); }
int main() { func_1(); }
\end{lstlisting}
\end{minipage}

\newparabold{Backtrace Invariant (BI)} 
The Backtrace Invariant is violated when the trace from optimized code contains a step over a line $l$ for which the stack backtrace includes a function name that is not present in the stack backtrace of any step in the trace from the unoptimized code that refers to the same line $l$.
The rationale behind BI is that the optimized code cannot reach a target function \cd{f} using a path on the call graph that is never used by the unoptimized code, since this would imply a latent violation of the LI invariant.

\newpara{Formal definition} Given two traces \emph{T(opt)} and \emph{T(unopt)}, the invariant is violated if $\exists s \in T(opt)$ and $\forall s' \in T(unopt)$ such that $l(s)=l(s')$ we have $BT(s) \not\subseteq BT(s')$.

\begin{wrapfigure}[17]{r}{0.22\textwidth}
	\vspace{-0.6cm}
\begin{lstlisting}[caption={SI violation. At line 2 variables  \cd{i,j,k}  are visibile, even if out of scope.},captionpos=b,style=CStyle,label=SI]
static int a[1];
int(b)(c, d) { return (c & (c ^ 7) - d) < 0 ? 0 : c - d; }
short(e)(short c) { return c; }
static int *f() {
  short g = 6;
  for (; g > 0; g = b(g, 2))
    e(g);
  *a = g;
  return a;
}
int main() {
  int i, j, k, print_hash_value;
  f();
  printf("%X\n");
}
\end{lstlisting}
\end{wrapfigure}

\newpara{Violation of BI} Snippet \ref{snippet1} shows C code that violates BI when compiled with clang and -$Og$. In this case, despite the absence of inlining in the optimized assembly, a step on line 8 shows an inconsistent backtrace. In our formalism, $\exists s \in T(opt)$ with $l(s)=8$ and $BT(s)=\{$\cd{main}, \cd{func\_1}, \cd{func\_2}$\}$, while for all steps on line $8$ in \emph{T(unopt)} the backtrace is $\{$\cd{main}$\}$.

\newparabold{Scope Invariant (SI)}
This invariant searches for lines of source code that are stepped over in both traces, but where there is at least one variable that is only visible in the trace from the optimized code.
Essentially, when SI is violated, there is a point in the optimized trace where an out-of-scope variable is visible. This behavior is not desirable during the debugging of an optimized binary as it would provide misleading information that does not match the visibility inferred by looking at the source code.

\newpara{Formal definition} Given traces \emph{T(opt)} and \emph{T(unopt)}, the invariant is violated if $\exists s \in T(opt)$ and $\exists s' \in T(unopt)$ with $l(s)=l(s')$ and $V(s) \not\subseteq V(s')$.

\newpara{Violation of SI} In Snippet \ref{SI}, there is a $C$ code that violates SI when compiled with GCC and -$Og$ and debugged using lldb. In this case, the third time we step on line $2$ the variables \cd{i,j,k} are visible even if they are declared in another scope.

\begin{wrapfigure}[14]{r}{0.22\textwidth}
\vspace{-0.5cm}
   \begin{lstlisting}[caption={PI violation, \cd{p\_6} seems to assume value $-1$.},captionpos=b,style=CStyle, label=snippet4]
typedef unsigned u32_t;
char c; u32_t d; int e;

int a(b) { return b; }
static int fun(u32_t p_6) {
  for (; c; c = 5)
    p_6 || d;
}
int main() {
  int f;
  e = a(8);
  f = e && 9;
  fun(f);
}
\end{lstlisting}
\end{wrapfigure}

\newparabold{Parameters Invariant (PI)}
With Parameters Invariant, we focus only on the variables that are also parameters of some function \cd{f} in the source code $C$. PI is violated if, in the trace from optimized code, there is a parameter with a value that is never observed in the trace from unoptimized code.
Intuitively, it exists a step where a parameter gets a value that is not consistent with the source code $C$ (a disagreement implies a bug, assuming that one of the traces is correct)\footnote{We do not have any source of randomness in our programs, that are deterministic. Moreover, we excluded pointers since the memory layout may change if some variables are optimized out (e.g., pointers to global strings, arrays, etc.)}

\newpara{Formal definition} Let $Par(P)$ be the set of all function parameters observed in trace $T(P)$, i.e. $Par(P):\{v | \exists s \in T(P) \land v \in V(s) \land v$ is parameter of a function \cd{f}$\}$; for each $v \in Par(P)$ let $Values(v,P)$ be the set of all values $v$ assumes in trace $T(P)$, i.e. $Values(v,P):\{val | \exists s \in T(P) \land v \in V(s) \land val=value(v,s) \}$. Then, PI is violated if $\exists v \in V(opt) \cap V(unopt)$ and $Values(v,opt) \not\subseteq Values(v,unopt) \cup \{\bot\}$.  We use the $\bot$ symbol to model the special value that is assigned by debuggers to optimized out variables.

\newpara{Violation of PI} Snippet \ref{snippet4} shows C code that violates PI when compiled with clang and -$O3$. In this case, at line 6 the parameter \cd{p\_6} is observed in the trace from optimized code having value $-1$ while its actual value is $1$; interestingly, this bug is also visible at line 14 of the same snippet where the value of variable \cd{f} in the same trace is $-1$.



\section{Implementation} \label{sec:imp}
%
%
We implemented \n for the  LLVM, GNU, and Rust toolchains.
We first detail the LLVM implementation and then highlight the differences for other toolchains.

\newpara{Program Generation}  In our default program generation module, we generate C source code using Csmith 2.4.0~\cite{yang2011finding} and use CompCert 3.7~\cite{Leroy-BKSPF-2016} to discard sources containing undefined behavior. 

\newpara{Tested toolchain} The toolchains under analysis are integrated in \optocus using Python. We automate lldb by using the Python bindings. Each trace is extracted by setting a breakpoint on the program entry point (\cd{break main}) and then by repeatedly stepping
(using the \cd{s} command).
At each step, we collect the corresponding source line number, function names in the stack trace using the \cd{bt} command, as well as variables and values returned by using the \cd{frame var} command.

\newpara{Trace consistency} We implemented this module using Python, following the design detailed in \S\ref{sec:generalidea}. The only variation is that, when checking for invariant PI, we discard all pointers because they are not guaranteed to be stable across executions (e.g., due to ASLR and non-deterministic heap allocators).

\newpara{Violations triaging} We implemented our fingerprinting mechanism using the opt-bisect tool~\cite{optbisect}, a tool provided by LLVM to bisect the optimization stages. Given C code that violates an invariant, we consider six versions of clang (i.e., 5, 6, 7, 8, 9, trunk), and on each of them, we use opt-bisect to pinpoint the optimization pass that generates the violation. 
We cluster together code samples that trigger violations with the same fingerprint and select a representative test case at random. We use C-Reduce 2.10~\cite{CReduce} to reduce the test case to a minimal source snippet triggering the same violation. We manually analyze the reduced test case to confirm that the violation is caused by a bug before reporting it to the developers.

\newparabold{Other toolchains}
The implementation for the GNU and Rust toolchains are similar to the LLVM one, but with the following notable exceptions. 
Since there seems to be no mature program generator for Rust, we simply collected a variety of source samples from the Rust test suite\footnote{\url{https://github.com/rust-lang/rust/tree/master/src/test}}. Our final dataset is composed by 12,616 source samples. We implemented the toolchain module for gdb using the pygdb library~\cite{pygdbmi}, which leverages the machine interface (gdb/mi) provided by gdb. We performed trace extraction similar to the LLVM toolchain using the specific commands for gdb. Finally, since the GNU and Rust toolchains have no equivalent of opt-bisect, in our experiments, we implemented the triaging procedure by means of manual analysis.

\section{Evaluation}\label{sec:eval}

Our experimental evaluation aims at answering the following main research questions:
 \begin{mdframed}[backgroundcolor=blue!8] 

\begin{description}
\item {\bf RQ 1:} {\em Is there a relationship between invariant violations and optimization levels? And between violations and actual bugs? }

\item {\bf RQ 2:} {\em Does \n find real bugs in our reference toolchain (LLVM) across different optimization levels and components (e.g., compiler and debugger)?}
\item {\bf RQ 3:} {\em Does \n generalize to different toolchains and programming languages?}

\end{description}
\end{mdframed}

We also report results that shed light on other interesting aspects: the relationship between optimization levels and certain bugs, the optimization passes that are more prone to bugs, and how our framework fares when swapping the program generation module.

To answer all the questions above, we evaluated \n on various target toolchains using the prototype described in \S\ref{sec:imp}. The main focus of our evaluation is the LLVM toolchain for the C language, namely the clang compiler and the lldb debugger. We also ran tests on the GNU toolchain (GCC and gdb) and on the Rust toolchain (rustc and lldb) aimed at investigating the generality of our framework among different toolchains and programming languages.
All our experiments run on Google Cloud VMs equipped with 32GB of RAM and 8 cores each; experiments run in Docker containers with Ubuntu 18.04.
  

%


\vspace{-0.1cm}
\subsection{Trace Invariant Violation Analysis} \label{sec:expviol}
\vspace{-0.1cm}

As a first step in our experimental evaluation, we want to analyze if and how much the trace invariants defined in \S\ref{sec:generalidea} are suitable to identify inconsistencies in the traces produced by the LLVM toolchain. Moreover, we want to study how these violations are distributed across optimization levels and if our fingerprinting mechanism can help reduce the manual effort by clustering violations sharing the same root cause.

We ran 24 hours of tests with optimization levels -$O[1,2,3,g,s,z]$ and stopped after collecting 7,500 test cases for each level. We tested each optimization level independently, i.e., the set of source samples generated by Csmith was different from level to level. We ran experiments using LLVM version 11.0.0 commit 80$\dots$f996977f. Table~\ref{table:violclangall} presents our results.

Looking at the raw number of violations (left side of the table), there are a number of interesting findings. First, the majority of LI violations happen at optimization levels \mbox{-$O[1,g]$}. This is probably because higher optimization levels more aggressively drop line information, hence limiting the amount of inconsistent information.
Interestingly, the opposite behavior can be observed when looking at PI violations: they reach their maximum number on optimization levels -$O[2,3]$, they are still high at -$O[s,z]$, but have a minimum at -$O[1,g]$. We speculate this behavior stems from inter-procedural optimizations and the use of inlining, passes that are disabled at -$O[1,g]$.
Considering BI, a large number of violations are at -$Og$. We manually analyzed such violations and found a pathological test case that causes most of them. As such, the large number is not statistically significant.


\begin{table}
\centering
\caption{Violations found for each optimization level in LLVM.}
\label{table:violclangall}
\begin{tabular}{c | cccc | cccc }
\toprule
{} & \multicolumn{4}{c}{\textbf{All Violations}} & \multicolumn{4}{c}{\textbf{Unique Fingerprints}} \\
Opt. &   LI & SI &  BI &   PI &               LI & SI & BI &  PI \\
\midrule
\textbf{01} &               54 &   3 &   1 &    2 &                10 &   3 &  1 &   2 \\
\textbf{02} &                3 &   6 &   2 &  114 &                 3 &   3 &  2 &  25 \\
\textbf{03} &                4 &   8 &   0 &  135 &                 2 &   2 &  0 &  27 \\
\textbf{0g} &               54 &   1 &  57 &    4 &                11 &   1 &  3 &   3 \\
\textbf{0s} &                5 &   2 &   0 &   52 &                 5 &   1 &  0 &  16 \\
\textbf{0z} &                3 &   3 &   2 &   78 &                 3 &   3 &  1 &  25 \\
\bottomrule
\end{tabular}
\end{table}

\color{black}

For each invariant, we then collected the unique fingerprints and discarded all duplicates (i.e., we considered a single representative test case per cluster).
Looking at the violations found for optimization level -$Og$, we have 11 unique fingerprints for the LI violations, 1 for SI, 3 for BI, and 3 for PI. However, the total number of unique fingerprints across all violations is 16 (the total is not reported in the table), i.e., less than the sum of unique fingerprints for each invariant since a single fingerprint may incur different invariant violations.

The large gap between the total number of violations and the number of unique fingerprints can be explained by the fact that a single bug may trigger several violations. For example, by analyzing the LI violations, we found that 41 are on a ternary operator; these are all caused by a single bug (see bug \aurl{https://bugs.llvm.org/show_bug.cgi?id=\bugid{45895}}) and they map to 5 different fingerprints.  For BI, we have 47 violations generated by the same bug on the same file, justifying only 3 unique fingerprints.
Regarding the PI violations, we found the same behavior mentioned above for optimization levels $O[2,3,s,z]$. In general, these results show that our framework, equipped with our four trace invariants, can identify a fairly large number of violations across several optimization levels.

\newpara{Relationship with bugs and fingerprints}
We ran an experiment to understand the number of violations caused by a bug. 
We started our analysis on April 3, 2020 with LLVM commit b73 $\ldots$ a7d2ed267b ({\em initial version}). We generated 1k test cases compiled using the initial version at -$Og$ and we found 1,956 violations. From April 3 to the end of June we reported several bugs (additional details in the next section), 14 of which have been patched. We then considered the LLVM version of June 26, 2020 (containing all the patches of our bugs) as the {\em final version}. We tested the 1k test cases again on the final version and found only 7 violations. The total breakdown (initial version / final version) by invariant is LI from 31 to 6, SI from 755 to 0, BI from 9 to 0, and PI from 1,161 to 1.
The total decrease from an average of $1.9$ violations per test case to $7 \cdot 10^{-3}$ shows that the vast majority of violations found in the initial version (over 99\%) was caused by a bug. This confirms that an invariant violation has a pathological cause most of the time.
During this experiment, we found that three bugs generate most violations: a first bug generates 1,144 violations, mapped to 17 fingerprints; a second bug generates 756 violations, mapped to 120 fingerprints; a third bug generates 41 violations, mapped to 11 fingerprints. This also shows that fingerprints effectively cluster violations from the same bug reducing the number of test case to analyze manually.

\subsection{Manually analyzed and reported bugs}\label{sec:bugs}

We present a quantitative analysis of the bugs we reported. A more qualitative analysis of selected bugs is available in \S\ref{sec:bugdiscussion}.

\smallskip\noindent\textbf{LLVM toolchain}\label{sec:llvmexp}

\noindent We collected several violations found by running \n on the latest LLVM release from April 3, 2020. We used optimization levels -O$[$0,1,2,3,s,z$]$, clustered violations using their fingerprints, extracted the representative test cases' source code from the clusters, and reduced them to smaller test cases with C-Reduce.

\begin{table}[]
	\center
	\caption{Breakdown by invariant of reported bugs for LLVM.}\label{table:reportedclang}
	\begin{tabular}{|l|l|l|l|l|}
		\hline
		Triggered Invariant & Unconfirmed & Confirmed & Patched & Total \\ \hline
		Lines  Invariant (LI) & 0 & 4 & 7 & 11 \\ \hline
		Backtrace Invariant (BI) & 0 & 1 & 3 & 4 \\ \hline
		Scope Invariant (SI) & 1 & 2 & 0 & 3 \\ \hline
		Parameters Invariant (PI) & 1 & 0 & 4 & 5 \\ \hline
	\end{tabular}
\end{table}

\newpara{Analysis of bugs found} In total, we reported 23 unique bugs. As expected, our system identified bugs in the full toolchain: 16 bugs found on clang, and 7 on lldb.
Table~\ref{table:reportedclang} presents a breakdown of the bugs by triggered invariant\footnote{Whenever a bug triggered multiple invariants, we selected only one.}. As shown in the table, the distribution follows the one we reported in \S\ref{sec:expviol}.

Regarding the distribution of bugs between compiler and debugger, for LI we discovered only bugs in clang, while for BI, SI, and PI we observed mixed results: SI exposed 2 bugs in clang and 1 in lldb; PI exposed 3 bugs in lldb and 1 on clang; BI exposed 1 bug in clang and 3 bugs in lldb.
This behavior stems from source-line stepping being mostly dependent on the information present in the DWARF line table, which is created entirely by the compiler, while the debugger is only responsible to parse such information. Since LI is the invariant that is more likely to find misaligned or wrong line information, most of LI violations come from compiler bugs.
PI and SI find bugs in both components of the toolchain since they both assess the ability of the compiler to generate a correct DWARF table (see bug \aurl{https://bugs.LLVM.org/show_bug.cgi?id=\bugid{45923}}) and the ability of the debugger to correctly parse it (see bug \aurl{https://bugs.LLVM.org/show_bug.cgi?id=\bugid{46181}}).
Interestingly, most of the BI bugs are in the debugger. This is sensible as the debugger does the majority of work to build a reliable stack trace using its unwinder. Nevertheless, we also found a BI bug in clang (see bug \aurl{https://bugs.llvm.org/show_bug.cgi?id=\bugid{45883}}).
Apart from the bugs above, we also reported 4 additional bugs that are duplicated (i.e., on the 27 bugs that we reported 23 had not been marked as duplicate). This low rate of duplicates is encouraging, as it shows that  violation triaging works well when it is possible to identify the optimization pass that triggers the violation.

\begin{figure}
\center
\includegraphics[scale=0.5]{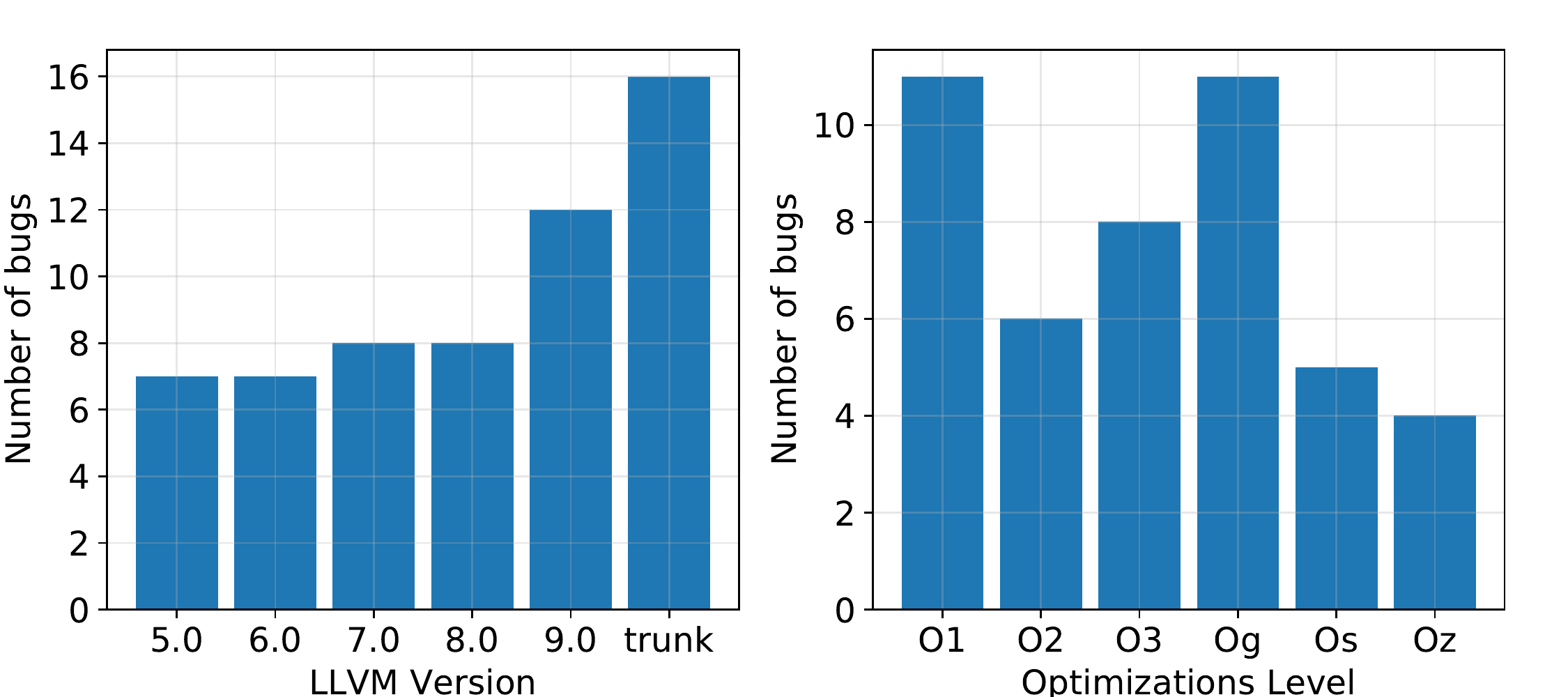}
\caption{Affected versions and optimization levels for reported bugs assigned to clang.}
\label{fig:LLVM_affected_versions}
\end{figure}

\newpara{Affected versions}
We tested for the presence of the 16 clang bugs on different versions of LLVM. Figure~\ref{fig:LLVM_affected_versions} presents a histogram of bugs per LLVM version. As shown in the figure, most of the bugs we found have been latent for years: for example, roughly half of the clang bugs also affect LLVM version 5.0, which was released in 2017.

\newpara{Affected optimization passes and levels}
As shown in Figure~\ref{fig:LLVM_affected_versions}, most of the bugs plague optimization levels -$Og$ and -$O1$. This can be explainable by the fact that those are the levels that drop less debug information.
From our analysis, we found that -$Og$ and -$O1$ share the same set of bugs (i.e., the 11 reported bugs are exactly the same) and the majority of these bugs are not present in -$O[2,3,s,z]$. This is not surprising since, after some investigation, we found that, in LLVM, -$Og$ is just an alias for -$O1$~\cite{clangmanual}. 
Table~\ref{table:optpass} presents a breakdown of the 16 clang bugs by the affected optimization pass. Interestingly, the more affected passes are the ones that either move instructions across the blocks of the CFG or that change the structure of the CFG. We speculate that is due to the difficulty of forecasting the effect of such transformations on debug information and due to the lack of specific guidelines on how to preserve debug information during these passes (more details are in \S\ref{sec:paperdiscussion}).

\begin{table}[]
\center

	\caption{LLVM bugs by affected optimization passes.}\label{table:optpass}
	\begin{tabular}{|l|l|}
		\hline
		Optimization Pass & \# bugs  \\ \hline
		Loop Strength Reduction & 1 \\ \hline
		Machine Common Subexpression Elimination & 1 \\ \hline
		Loop Invariant Code Motion & 2 \\ \hline
		CodeGen Prepare & 1 \\ \hline
		Stack Slot Coloring & 1 \\ \hline
		Control Flow Optimizer & 2 \\ \hline
		Simplify the CFG & 3 \\ \hline
		X86 DAG -\textgreater DAG Instruction Selection & 1 \\ \hline
		Branch Probability Basic Block Placement & 3 \\ \hline
		Rotate Loops & 1 \\ \hline
	\end{tabular}
\end{table}

\begin{table}[]
	\center
	\caption{Breakdown by invariant of reported bugs for GNU.}\label{table:reportedgcc}
	\begin{tabular}{|l|l|l|l|l|}
		\hline
		Triggered Invariant & Unconfirmed & Confirmed & Patched & Total \\ \hline
		Lines  Invariant (LI) & 4 & 1 & 0 & 5 \\ \hline
		Backtrace Invariant (BI) & 1 & 1 & 0 & 2 \\ \hline
		Scope Invariant (SI) & 0 & 0 & 0 & 0 \\ \hline
		Parameters Invariant (PI) & 1 & 0 & 0 & 1 \\ \hline
	\end{tabular}
\end{table}

\smallskip\noindent\textbf{Other toolchains}

\noindent Our analysis on LLVM has answered {\bf RQ 2}. In this section, we report on the experiments we ran on other toolchains to answer {\bf RQ 3}.

\newpara{GNU toolchain}
We performed the same experiments on the GNU toolchain (GCC and gdb). For this toolchain, we reported a total of 8 bugs, 2 of which have been confirmed (Table~\ref{table:reportedgcc}). Interestingly, we could not found a bug that generates a SI.
We tested the bugs on versions 5, 6, 7, 8, 9, 10, trunk, and found that the bugs affect only version 8 and more recent versions. This is probably due to a major change in the generation of DWARF debug information. 
On the trunk version, the breakdown by optimization level is 6 bugs for -$Og$ and 4 bugs for -$O[1,2,3,fast]$.
This confirms our hypothesis that optimization levels preserving more debug information are more vulnerable. Interestingly, we found the same set of bugs for levels -$O[2,3,fast]$ even if the optimization strategy differs among these levels. 


\newpara{Rust toolchain}
We used the latest rustc nightly build (1.46.09) for the compiler and lldb as debugger.
We compiled each sample in our dataset with the -$g$ flag and optimization levels -$O[1,3]$. We used each optimized binary as input to our trace consistency subsystem along with the non-optimized one. We manually analyzed a selected set of violations, identifying and reporting 3 bugs from 3 invariant violations: 1 LI, 1 SI, and 1 PI.


\subsection{Other program generators}\label{sec:codegeneval}
As noted earlier, the program generation component of \n is interchangeable by design. We show how changing the program generation component impacts the number of violations we can find, while at the same time confirming our \n framework is extensible by swapping components.

To this end, we ran the whole pipeline again (as described in
\S\ref{sec:eval}), but replaced
the default Csmith generator with 3 alternative approaches to program
generation. We also targeted a more recent clang/LLVM version (commit
94e4e37d5564), in which numerous bugs that we reported had been fixed.

Table~\ref{tab:geneval} shows the median number of unique fingerprints
and violations for 20 runs of 12 hours per code generation tool. We highlight
statistically significant values (Mann-Whitney $p < 0.05$), as compared to the
Csmith baseline. As shows in the table, yarpgen is capable of finding a
large number of new invariant violations not previously found by \n using
Csmith as the default program generation component. We also show how a simple
code mutation-based approach based on MUSIC~\cite{phan2018music} fares against
the generation-based approach taken by Csmith and yarpgen. The mutation-based
approach applies a number of (randomly selected) mutation operators on 5,000
initial seeds generated by Csmith and yarpgen. While our results suggest that a
generation-based approach outperforms this simple mutation-based approach, this
experiment does show \n generalizes to different classes of program generators.

\begin{table}[]
 \center
\begin{tabular}{l|rr|rr}
  & \multicolumn{4}{c}{\textbf{Unique fingerprints (total violations)}} \\ \hline
  &\multicolumn{2}{c}{\textbf{\Og}}  & \multicolumn{2}{c}{\textbf{\Otre}}  \\ \hline
  \textbf{Csmith}        & 4 & (5)   & 13 & (20)  \\
  \textbf{yarpgen}       & \textbf{11} & (\textbf{264}) & \textbf{14} & (\textbf{349}) \\
  \textbf{MUSIC-Csmith}  & \textbf{2} & (5) & \textbf{8}  & (\textbf{13})  \\
  \textbf{MUSIC-yarpgen} & \textbf{11} & (\textbf{251}) & 13 & (\textbf{331})
\end{tabular}
  \caption{Median number of fingerprints found per tool (N=20).}
 \label{tab:geneval}
\end{table}

\section{Detailed analysis of discovered bugs}\label{sec:bugdiscussion}
In this section, we detail some of the most interesting bugs we reported.

\begin{figure}

  \begin{lstlisting}[caption={lldb bug \bugid{46398} },captionpos=b,style=CStyle,label=46398]
int k = 0, a;
int *b = &a;
int c = 2;
short d = 2;
void e() { --d; }
int *f() {
  if (a)
    e();
  else
    return b;
}
int main() {
  f();
  f();
}
\end{lstlisting}

\end{figure}

\paragraph{lldb bug \bugid{46398} (Snippet \ref{46398})} 
At line 10, lldb shows two incoherent backtraces when the analyzed binary is compiled with -$Og$ and -$O0$. Specifically, it shows that variable \code{k} is a function in the stack trace.

The unwinder is the subsystem in lldb responsible for reconstructing the stack chains when stopped at a given address. The debugger implements unwinding as a series of plans which kick in one after another until one is successful (or the stack cannot be reconstructed, throwing an error). The wrong backtrace was caused by a bug in the unwind choosing strategy, which did not take into account the availability of additional information to pick the best strategy.


\paragraph{lldb bug \bugid{46456} (Snippet \ref{46456})} presents a bug in the DWARF parsing and interpretation logic of lldb. This bug was discovered performing a  {\em ``mix and match''} testing using GCC as the compiler and lldb as the debugger. lldb shows a flow where it incorrectly appears that the function \code{b()} is called twice. 

\begin{wrapfigure}[6]{r}{0.24\textwidth}
 \vspace{-0.6cm}
  \begin{lstlisting}[caption={lldb bug \bugid{46456}},captionpos=b,style=CStyle,label=46456]
int a;
void b() { printf("%X\n"); }
int main() {
  a = 1;
  b();
}
\end{lstlisting}
\end{wrapfigure}

The degradation is caused by the debugger not correctly parsing a DWARF attribute, namely \code{DW\_LNS\_negate\_stmt}. This class of bugs justifies and motivates the importance of performing further mixed consumer/producer testing. As the set of people working on lldb and clang are sometimes overlapping, as well as the set of people working on GCC and GDB, having multiple implementations of the standard tested reveals holes like the one outlined above.
 \begin{figure}
   \begin{lstlisting}[caption={clang bug \bugid{46008}},captionpos=b,style=CStyle,label=46008]
short a = 1;
int b, c;
void(d)() {}
int main() {
  int *p_5 = &c;
  for (; a <= 0; a = d)
    if ((*p_5))
      break;
  b = *p_5;
}
\end{lstlisting}
\end{figure}
 \paragraph{clang bug \bugid{46008} (Snippet \ref{46008})} 
 When stepping through optimized code, lldb steps on a source line that is dead (cannot be executed). In this case it shows the execution of line 7. This is due to the \emph{SimplifyCFG} optimization pass in LLVM, which performs peephole optimizations of the control flow graph, like flattening or block merging, sometimes without proper updates to line information while applying the transformations. In this specific case, the pass tries to fold a dead block into a live one, wrongly dropping the location. As a result, the debug information points to a ``dead'' line.

 \begin{figure}
   \begin{lstlisting}[caption={GCC bug \bugid{95077}},captionpos=b,style=CStyle,label=95077]
static int a, c, d, e;
static void dm() {
  int b = 0;
  for (; b < 56; b++)
    a = b;
}
int main() {
  int k;
  dm();
  d = 0;
  for (; d != 38; d = d + 1)
    e = 0;
  for (; e < 3; e++)
    for (k = 0; k < 4; k++)
      printf("", c);
}
\end{lstlisting}
\end{figure}

\paragraph{GCC bug \bugid{95077}  (Snippet \ref{95077})}  
In this case, at line 14 GDB shows that the top frame of the stack trace is function \code{dm}, despite line 14 being in the main. This bug stems from how GCC handles debug information of inlined functions. 
During the optimization passes, function \code{dm} is inlined; the mapping between inlined assembly instructions and the \code{dm} symbol is kept by DWARF using a table that associates ranges of assembly instructions to inlined functions.

 In this case, GCC wrongly includes an assembly instruction of the loop at line 14 in such ranges.  Therefore, when GDB parses this information, it wrongly shows that line 14  belongs to inlined function \code{dm}.

\section{Discussion}\label{sec:paperdiscussion}
\vspace{-0.1cm}
In this sections, we discuss the insights that we learned during our study, as well as the limitations of our work.

\medskip\noindent\textbf{The need for shared guidelines}\label{sec:guidelines}

\noindent During our investigation (see \S\ref{sec:bugs} and \S\ref{sec:bugdiscussion}) we reported several bugs that lead to interesting discussions among the toolchains developers.
During this process we realized that there has never been an effort to precisely define how to preserve
debug information during the optimization passes of production compilers (we are not aware of guidelines in this sense).
Therefore, several of our bugs required non-trivial fixes decided after lengthy discussions among toolchains developers.

A prototypical example is the bug reported in the Snippet \ref{spec}, where the code compiled with clang and optimization level -$Oz$ shows a step on line 2.  If we look at the source code, it is apparent that line 2 is never executed; and so it is in the unoptimized program.
However, in the optimized binary, due to a loop rotation optimization kicking in \cite{optimizingbook}, line 2 is indeed speculatively executed before reaching the end of the loop.
 One could argue either against or in favor of showing the step at line 2.

\begin{lstlisting}[caption={LSI violation in clang  when compiled with -$Oz$.  Line 2 appears to be executed.}% Bug report: \url{https://bugs.llvm.org/show_bug.cgi?id=\bugid{45934}}
,captionpos=b,style=CStyle,label=spec]
int add(int ui1, int ui2) {
  return ui1 + ui2;
}
int g_8 ;
int a = 4 ;
int c;
int main()  {
  for (g_8 = 0; (g_8 < 49); g_8 = add(g_8, 9)) {
      for (; c ; c++)
        ;
    if (a)
      break;
  }
}
\end{lstlisting}

On the one hand, one could follow the general principle of optimization transparency pioneered by Hennessy \cite{hennessy1982symbolic}: showing the step could mislead a developer, that would see a step over a location that should not be executed, this could be extremely confusing on large and complex software.
On the other hand, it could be reasonable to show what is precisely happening without masking the optimization (following the line of Brooks et al. \cite{brooks1992new}):  line 2 is executed even if this is done speculatively and without observable side effects.


In this specific case, clang developers finally decided that such behavior is indeed a bug and that the step at line 2 should have not been shown.

Similar design choices should derive from a broader discussion on the semantics of optimized debug information, since case by case fixes, in the long term, are likely to lead to inconsistent behaviors.
Hence, our investigation is a precursor of a more in-depth process that should systematically study how to preserve debug information during optimization passes, taking specifically into account standards for debug information (such as DWARF as we will discuss in the next section). 

During the production of this paper, the LLVM developers published a document containing a set of basic rules to update debug information during certain optimization passes \cite{vedantkumarDBG}. The content of this document has clearly been influenced by the bugs we reported, and it is a first step in the direction of defining shared guidelines.

\medskip\noindent\textbf{DWARF shortcomings}

\vspace{-0.01cm}

\noindent During our analysis, we encountered some bugs that highlighted a shortcoming of the DWARF standard that severely limits its expressivity, and impacts either the correctness or the completeness of the debug experience. In the following, we will show two examples, one for GCC and the other for clang, showcasing this problem. 
\begin{figure}
  \begin{lstlisting}[caption={GCC bug \bugid{95865}},captionpos=b,style=CStyle,label=95865]
int g_4 = 3, a;
int *b() {
  if (g_4)
    return &g_4; 
  a = 0;
  return &g_4; 
}
int main() { b(); }
\end{lstlisting}
\end{figure}

\paragraph{ GCC bug \bugid{95865} (Snippet \ref{95865})}
In this example GDB steps on dead code: it executes line 6 instead of 4. 
 The culprit is an optimization pass that modifies the CFG; 
 in this case line 4 and line 6 are lowered into a single assembly instruction while the instruction \code{a=0} becomes guarded by the condition \code{g\_4==0}.  In DWARF, each assembly instruction is mapped to at most single line of source code; hence, this case imposes a choice: the compiler may map the instruction to line 4 or line 6 (in this case a bug arises depending on the value of \code{g\_4}), or completely drop the source code location on the return (impacting the amount of information shown). Therefore, we have an unavoidable tradeoff between correctness (dropping the line) and completeness (picking a line and accepting a bug in some runs).

\begin{wrapfigure}[12]{r}{0.24\textwidth}
\vspace{-0.7cm}
   \begin{lstlisting}[caption={clang bug \bugid{45523}},captionpos=b,style=CStyle,label=45523]
volatile int a, b, c;
int g_3[2];
int main() {
  for (; b > -9; b--)
    ;
  for (; c <= 5; c++) {
    g_3[1] = 0;
    if (b)
      ;
    else {
      char l_1[3][4][3]={0};
      g_3[1] = l_1[2][3][2];
    }
  }
}
\end{lstlisting}
\end{wrapfigure}

\paragraph{ clang bug \bugid{45523} (Snippet \ref{45523})} shows a case where lldb hits the body of a dead branch, stepping on line 12 despite the condition in the \code{if} statement is true. The loop invariant code motion pass in LLVM tries to sink load/store instructions outside of the loop, and when there are several that have different locations, just picks an arbitrary one (in general, incorrect).

This bug was fixed collecting all the locations, and merging them in a single one, to be shown in the debugger, as long as they agree. In case there are different locations, the final location of the sunk instructions is dropped. For the reasons explained in the previous bug, this fix is inherently an approximation of the best behavior.

A possible future proposal for DWARF is to allow a single instruction to be mapped to multiple locations, letting the consumers (e.g., the debugger) decide the policy to adopt.

\section{Limitations, and threats to validity}


\noindent In our system, we have to generate programs that are free from undefined behaviors to avoid spurious violations of an invariant. 
The presence of undefined behavior could result in a non-deterministic program artificially inflating the number of violations. 
We use CompCert to filter programs with undefined behavior. 
While this is not provably perfect and does not rule out the possibility that undefined behavior is still present, we have never observed one during our tests.

Whenever a violation is found, we must manually check if the root cause is a bug. While \n does not yet provide full automatic triaging, we found that our triaging module significantly speeds up the analysis of violations. 

Similarly to all previous works using differential testing, \n cannot detect a bug that is present in both analyzed traces. This issue could be solved by designing invariants that work on a single trace, based on  assumptions on the code behavior to detect inconsistencies. Future work could investigate this aspect.

\section{Related Work}\label{sec:relatedwork}

The investigation on the correctness of debug information started with the work of Yuanbo Li et al.
 \cite{davidepldi}. 
  The paper focuses on validating the correctness of the values of the variables shown by a debugger for optimized code for statically compiled languages. Their idea is to derive from a certain source code a modified program, built with optimizations, in which a debugger can stop at a predetermined line and print the value of a specific variable without triggering undefined behavior. 
Once stopped, they compare this value to the one printed, at the same program point, by an unoptimized binary.

Our work generalizes their:  they are only looking at one aspect of debug information, consistency of variables information, while they neglect other aspects (e.g., they assume line information to be correct).
Apart from \cite{davidepldi}, the majority of the other works in this area look at loosely related problems, and they can mainly be partitioned in the works on improving the experience of debugging optimized binaries, the ones that focus on testing the correctness of debuggers, and the massive amount of work devoted to compilers testing.

\paragraph{Debugging of optimized binaries}

Generating debug information 
for optimized binaries is a long-studied problem. The seminal work of Hennessy \cite{hennessy1982symbolic} proposed a categorization of the effect of optimizations on the variables of a program. It also introduced an algorithm to identify endangered variables (variables whose values are possibly not correct in the optimized program). The main idea behind his approach was to make optimizations transparent and provide the user with the expected behavior. Building on top of \cite{hennessy1982symbolic}, \cite{wismuller1994debugging,adl1996source} proposed better approaches for identifying endangered variables.


Conversely, from \cite{hennessy1982symbolic}, Brooks et al. \cite{brooks1992new} proposed an alternative principle to follow: try to depict what is happening in the optimized code, without striving for the illusion of transparency. To reach this goal, they aim to provide the user with visual-feedback highlighting pieces of source code during interactive debugging. 
Finally, to identify possible bugs in optimized code, Jaramillo et al. \cite{jaramillo1999comparison}, proposed comparison-checking between unoptimized and optimized code after executing them under the same input. We remark that they are interested in finding miscompiles introduced by the optimization passes, and not in wrong debug information.

\paragraph{Testing of debuggers}
Several works investigate the problem of finding debugger bugs. A differential testing approach was taken by Lehmann et al. \cite{lehmann2018feedback}; they proposed to record debug traces for multiple debuggers and compare them to identify bugs. Such an approach can be used only when multiple mature debuggers are available. Considering a non-differential approach, Tolksdorf et al. \cite{tolksdorf2019interactive} showed how it is possible to identify bugs in Chromium debugger using metamorphic testing. We point out that none of the aforementioned works is  able to find bugs in the compiler.

\paragraph{Compilers testing}
The problem of compiler testing received a lot of attention in the past decade \cite{yang2011finding,le2014compiler,sun2016toward} and it mainly focused on finding miscompiles in optimizing compilers. For a recent survey on  compiler testing techniques, see Chen et al.  \cite{chen2020survey}. This research line is complementary to our since it is devoted to test the correctness of machine code generated by the compiler without caring about debug information. 

\section{Conclusions}\label{sec:conclusions}

This paper introduced \n, a framework to expose debug information bugs in production toolchains. \n fills an important gap in the literature, where little attention has been devoted to scrutinizing the full debug information lifecycle for bugs. This is particularly problematic when debugging happens on production software, where binaries have been heavily optimized, and debug information bugs may hinder post-deployment debugging efforts.
\n relies on trace invariants to perform differential analysis on debug traces of optimized and unoptimized programs, and expose inconsistencies whose root cause may be a bug in the compiler or in the debugger.
We evaluated \n on three different toolchains (namely LLVM, GNU, and Rust), finding 34 new bugs. Most bugs have already been fixed by the developers. Furthermore, our findings have already sparked an interesting discussion among developers and fostered further investigation on the semantics of debug information for optimized programs.

\paragraph{Acknowledgment:}

The VU Amsterdam researchers were supported by the European Union's Horizon 2020 research and innovation programme under grant agreement No. 786669 (ReAct) and by Cisco Systems, Inc. through grant \#1138109.
Giuseppe Di Luna was supported by the AXA Postdoctoral Fellowship.  We thank Daniele Cono D'Elia for his comments on a previous version of this manuscript.

\bibliographystyle{plain}
\bibliography{references}

\begin{thebibliography}{10}

\bibitem{clangmanual}
Clang 12 documentation.
\newblock \url{https://clang.llvm.org/docs/CommandGuide/clang.html}, 2020.
\newblock [Online; accessed 27-July-2020].

\bibitem{optbisect}
Using -opt-bisect-limit to debug optimization errors.
\newblock \url{https://llvm.org/docs/OptBisect.html}, 2020.
\newblock [Online; accessed 27-July-2020].

\bibitem{adl1996source}
Ali-Reza Adl-Tabatabai and Thomas Gross.
\newblock Source-level debugging of scalar optimized code.
\newblock In {\em Proceedings of the ACM SIGPLAN 1996 conference on Programming
  language design and implementation}, pages 33--43, 1996.

\bibitem{yarpgen}
Dmity Babokin, John Regehr, and Vsevolod Livinskiy.
\newblock Yarpgen: Yet another random program generator.
\newblock \url{https://github.com/intel/yarpgen}, 2020.
\newblock [Online; accessed 27-July-2020].

\bibitem{brooks1992new}
Gary Brooks, Gilbert~J Hansen, and Steve Simmons.
\newblock A new approach to debugging optimized code.
\newblock {\em ACM SIGPLAN Notices}, 27(7):1--11, 1992.

\bibitem{chen2020survey}
Junjie Chen, Jibesh Patra, Michael Pradel, Yingfei Xiong, Hongyu Zhang, Dan
  Hao, and Lu~Zhang.
\newblock A survey of compiler testing.
\newblock {\em ACM Computing Surveys (CSUR)}, 53(1):1--36, 2020.

\bibitem{copperman1994debugging}
Max Copperman.
\newblock Debugging optimized code without being misled.
\newblock {\em ACM Transactions on Programming Languages and Systems (TOPLAS)},
  16(3):387--427, 1994.

\bibitem{7163211}
V.~{D'Silva}, M.~{Payer}, and D.~{Song}.
\newblock The correctness-security gap in compiler optimization.
\newblock In {\em 2015 IEEE Security and Privacy Workshops}, pages 73--87,
  2015.

\bibitem{hennessy1982symbolic}
John Hennessy.
\newblock Symbolic debugging of optimized code.
\newblock {\em ACM Transactions on Programming Languages and Systems (TOPLAS)},
  4(3):323--344, 1982.

\bibitem{jaramillo1999comparison}
Clara Jaramillo, Rajiv Gupta, and Mary~Lou Soffa.
\newblock Comparison checking: An approach to avoid debugging of optimized
  code.
\newblock In {\em Software Engineering---ESEC/FSE'99}, pages 268--284.
  Springer, 1999.

\bibitem{6595791}
C.~{Jia} and W.~K. {Chan}.
\newblock Which compiler optimization options should i use for detecting data
  races in multithreaded programs?
\newblock In {\em 2013 8th International Workshop on Automation of Software
  Test (AST)}, pages 53--56, 2013.

\bibitem{optimizingbook}
Ken Kennedy and John~R. Allen.
\newblock {\em Optimizing Compilers for Modern Architectures: A
  Dependence-Based Approach}.
\newblock Morgan Kaufmann Publishers Inc., San Francisco, CA, USA, 2001.

\bibitem{vedantkumarDBG}
Vedant Kumar.
\newblock How to update debug info: A guide for llvm pass authors.
\newblock
  \url{https://github.com/llvm/llvm-project/blob/master/llvm/docs/HowToUpdateDebugInfo.rst},
  2020.
\newblock [Online; accessed 27-July-2020].

\bibitem{le2014compiler}
Vu~Le, Mehrdad Afshari, and Zhendong Su.
\newblock Compiler validation via equivalence modulo inputs.
\newblock {\em ACM SIGPLAN Notices}, 49(6):216--226, 2014.

\bibitem{lehmann2018feedback}
Daniel Lehmann and Michael Pradel.
\newblock Feedback-directed differential testing of interactive debuggers.
\newblock In {\em Proceedings of the 2018 26th ACM Joint Meeting on European
  Software Engineering Conference and Symposium on the Foundations of Software
  Engineering}, pages 610--620, 2018.

\bibitem{Leroy-BKSPF-2016}
Xavier Leroy, Sandrine Blazy, Daniel K\"astner, Bernhard Schommer, Markus
  Pister, and Christian Ferdinand.
\newblock Compcert -- a formally verified optimizing compiler.
\newblock In {\em ERTS 2016: Embedded Real Time Software and Systems}. SEE,
  2016.

\bibitem{davidepldi}
Yuanbo Li, Shuo Ding, Qirun Zhang, and Davide Italiano.
\newblock Debug information validation for optimized code.
\newblock In {\em PLDI}, pages 1052--1065, 2020.

\bibitem{phan2018music}
Duy~Loc Phan, Yunho Kim, and Moonzoo Kim.
\newblock Music: Mutation analysis tool with high configurability and
  extensibility.
\newblock In {\em 2018 IEEE International Conference on Software Testing,
  Verification and Validation Workshops (ICSTW)}, pages 40--46. IEEE, 2018.

\bibitem{CReduce}
John Regehr, Yang Chen, Pascal Cuoq, Eric Eide, Chucky Ellison, and Xuejun
  Yang.
\newblock Test-case reduction for c compiler bugs.
\newblock In {\em Proceedings of the 33rd ACM SIGPLAN Conference on Programming
  Language Design and Implementation}, PLDI '12, pages 335--346, New York, NY,
  USA, 2012. Association for Computing Machinery.

\bibitem{schulte2014post}
Eric Schulte, Jonathan Dorn, Stephen Harding, Stephanie Forrest, and Westley
  Weimer.
\newblock Post-compiler software optimization for reducing energy.
\newblock {\em ACM SIGARCH Computer Architecture News}, 42(1):639--652, 2014.

\bibitem{pygdbmi}
Chad Smith.
\newblock pygdbmi - get structured output from gdb's machine interface.
\newblock \url{https://github.com/cs01/pygdbmi}, 2020.
\newblock [Online; accessed 27-July-2020].

\bibitem{sun2016toward}
Chengnian Sun, Vu~Le, Qirun Zhang, and Zhendong Su.
\newblock Toward understanding compiler bugs in gcc and llvm.
\newblock In {\em Proceedings of the 25th International Symposium on Software
  Testing and Analysis}, pages 294--305, 2016.

\bibitem{tolksdorf2019interactive}
Sandro Tolksdorf, Daniel Lehmann, and Michael Pradel.
\newblock Interactive metamorphic testing of debuggers.
\newblock In {\em Proceedings of the 28th ACM SIGSOFT International Symposium
  on Software Testing and Analysis}, pages 273--283, 2019.

\bibitem{wismuller1994debugging}
Roland Wism{\"u}ller.
\newblock Debugging of globally optimized programs using data flow analysis.
\newblock In {\em Proceedings of the ACM SIGPLAN 1994 conference on Programming
  language design and implementation}, pages 278--289, 1994.

\bibitem{yang2011finding}
Xuejun Yang, Yang Chen, Eric Eide, and John Regehr.
\newblock Finding and understanding bugs in c compilers.
\newblock In {\em Proceedings of the 32nd ACM SIGPLAN conference on Programming
  language design and implementation}, pages 283--294, 2011.

\bibitem{YIN201918}
Jie Yin, Gang Tan, Hao Li, Xiaolong Bai, Yu-Ping Wang, and Shi-Min Hu.
\newblock Debugopt: Debugging fully optimized natively compiled programs using
  multistage instrumentation.
\newblock {\em Science of Computer Programming}, 169:18 -- 32, 2019.

\end{thebibliography}

\end{document}